\documentclass[preprint2]{aastex631}

\graphicspath{{./}{figures/}}
\accepted{Feb 15, 2023}

\submitjournal{PASP}

\begin{document}

\title{Spectroscopic time series performance of the Mid-Infrared Instrument on the JWST}

\author[0000-0003-4757-2500]{Jeroen Bouwman}
\affiliation{Max Planck Institute for Astronomy (MPIA), K\"{o}nigstuhl 17, D-69117 Heidelberg, Germany}

\author[0000-0002-7612-0469]{Sarah Kendrew}
\affiliation{European Space Agency, Space Telescope Science Institute, 3700 San Martin Dr., Baltimore, MD 21218, USA}

\author[0000-0002-8963-8056]{Thomas P. Greene}
\affiliation{NASA Ames Research Center, Space Science and Astrobiology Division, MS 245-6, Moffett Field, CA, 94035, USA}

\author[0000-0003-4177-2149]{Taylor J.~Bell}
\affiliation{BAER Institute, NASA Ames Research Center, Moffet Field, CA 94035, USA}

\author{Pierre-Olivier Lagage}
\affiliation{Universit\'{e} Paris-Saclay, Universit\'{e} de Paris, CEA, CNRS, AIM, F-91191 Gif-sur-Yvette, France}

\author{J\"{u}rgen Schreiber}
\affiliation{Max Planck Institute for Astronomy (MPIA), K\"{o}nigstuhl 17, D-69117 Heidelberg, Germany}

\author{Daniel Dicken}
\affiliation{UK Astronomy Technology Centre, Royal Observatory Edinburgh, Blackford Hill, Edinburgh EH9 3HJ, UK}

\author{G.~C.~Sloan}
\affiliation{Space Telescope Science Institute, 3700 San Martin Dr., Baltimore, MD 21218, USA}
\affiliation{Dept.\ of Physics and Astronomy, University of North Carolina at Chapel Hill, Campus Box 3255, Chapel Hill, NC 27599, USA}

\author[0000-0001-9513-1449]{N\'estor Espinoza}
\affiliation{Space Telescope Science Institute, 3700 San Martin Dr., Baltimore, MD 21218, USA}

\author[0000-0003-4559-0721]{Silvia Scheithauer}
\affiliation{Max Planck Institute for Astronomy (MPIA), K\"{o}nigstuhl 17, D-69117 Heidelberg, Germany}

\author[0000-0001-6492-7719]{Alain Coulais}
\affiliation{LERMA, Observatoire de Paris, Universit\'{e} PSL, Sorbonne Universit\'{e}, CNRS, 
Paris, France}
\affiliation{AIM, CEA, CNRS, Universit\'{e} Paris-Saclay, Universit\'{e} de Paris, F-91191 Gif-sur-Yvette, France}

\author{Ori D. Fox}
\affiliation{Space Telescope Science Institute, 3700 San Martin Dr., Baltimore, MD 21218, USA}

\author{Ren\'{e} Gastaud}
\affiliation{AIM, CEA, CNRS, Universit\'{e} Paris-Saclay, Universit\'{e} de Paris, F-91191 Gif-sur-Yvette, France}

\author{Adrian M. Glauser}
\affiliation{ETH Z\"{u}rich, Wolfgang-Pauli-Str. 27, 8093 Z\"{u}rich, Switzerland}

\author{Olivia C. Jones}
\affiliation{UK Astronomy Technology Centre, Royal Observatory Edinburgh, Blackford Hill, Edinburgh EH9 3HJ, UK}

\author[0000-0002-0690-8824]{Alvaro Labiano}
\affiliation{Telespazio UK for the European Space Agency, ESAC, Camino Bajo del Castillo s/n, 28692 Villanueva de la Ca\~nada, Spain.}
\affiliation{Centro de Astrobiolog\'ia (CSIC-INTA), Carretera de Ajalvir, 28850 Torrej\'on de Ardoz, Madrid, Spain.}

\author{Fred Lahuis}
\affiliation{SRON Netherlands Institute for Space Research, PO Box 800, 9700 AV, Groningen, The Netherlands}

\author{Jane E. Morrison}
\affiliation{Steward Observatory,University of Arizona, Tucson, AZ, 85721, USA}

\author{Katherine Murray}
\affiliation{Space Telescope Science Institute, 3700 San Martin Dr., Baltimore, MD 21218, USA}

\author[0000-0003-3217-5385]{Michael Mueller}
\affiliation{Kapteyn Astronomical Institute, Rijksuniversiteit Groningen, Postbus 800, 9700AV Groningen, The Netherlands}

\author[0000-0001-6576-6339]{Omnarayani Nayak}
\affiliation{Space Telescope Science Institute, 3700 San Martin Dr., Baltimore, MD 21218, USA}

\author{Gillian S. Wright}
\affiliation{UK Astronomy Technology Centre, Royal Observatory Edinburgh, Blackford Hill, Edinburgh EH9 3HJ, UK}

\author{Alistair Glasse}
\affiliation{UK Astronomy Technology Centre, Royal Observatory Edinburgh, Blackford Hill, Edinburgh EH9 3HJ, UK}

\author{George Rieke}
\affiliation{Steward Observatory,University of Arizona, Tucson, AZ, 85721, USA}



\begin{abstract}
We present here the first ever mid-infrared spectroscopic time series observation of the transiting exoplanet \object{L 168-9 b} with the Mid-Infrared Instrument (MIRI) on the James Webb Space Telescope. The data were obtained as part of the MIRI commissioning activities, to characterize the performance of the Low Resolution Spectroscopy (LRS) mode for these challenging observations. To assess the MIRI LRS performance, we performed two independent analyses of the data. We find that with a single transit observation we reached a spectro-photometric precision of $\sim$50 ppm in the 7-8~\micron~range at R=50, consistent with $\sim$25 ppm systematic noise. The derived band averaged transit depth is 524~$\pm$~15~ppm and 547~$\pm$~13~ppm for the two applied analysis methods, respectively, recovering the known transit depth to within 1~$\sigma$. The measured noise in the planet's transmission spectrum is approximately 15-20\% higher than random noise simulations over wavelengths $6.8 \lesssim \lambda \lesssim 11$ $\mu$m. \added{We observed an larger excess noise at the shortest wavelengths of up to a factor of two, for which possible causes are discussed.} This performance was achieved with limited in-flight calibration data, demonstrating the future potential of MIRI for the characterization of exoplanet atmospheres.

\end{abstract}



\section{Introduction} \label{sec:intro}

\subsection{Characterization of transiting exoplanet atmospheres with JWST}

A substantial fraction of the cycle 1 time of the newly launched James Webb Space Telescope \citep[JWST;][]{2006SSRv..123..485G} has been allocated to observations of exoplanets, for which the observatory offers numerous observational modes. A major area of study is the characterization of exoplanet atmospheres via transit spectroscopy. In these observations, the host star and its planet are observed throughout the planet's transit in front of the star (\textit{transmission spectroscopy}); the planet's passage behind the star (\textit{eclipse spectroscopy}); or for an entire orbital period (\textit{phase curve}). Observations with the Hubble Space Telescope (HST) and Spitzer Space Telescope (SST) have proven that such observations can provide valuable information on the chemical composition of exoplanetary atmospheres, and the presence of clouds and hazes.~\citet{madhu_review2019} gives a comprehensive review of the state of the art in transit spectroscopy, and the detections and insights these techniques have produced. 

\noindent JWST's broad infrared wavelength coverage and suite of capabilities promises transformative discoveries in exoplanet science. To provide optimal operational conditions, transit photometry or spectroscopy observations can be flagged as ``time series observations'' (TSO) in the Astronomers' Proposal Tool (APT). In JWST TSOs, observations are executed as single exposures of many integrations, whose duration sets the cadence of the time series. The limit on the duration of a single exposure is set by spacecraft limitations such as the maximum number of groups, integrations, or maximum total data volume that can be held in memory.

\noindent To enable TSOs over many hours, the 10,000-second limit on duration of a single exposure is waived, and the High Gain Antenna (HGA) is allowed to re-point while data is being recorded. Dithering is disabled to ensure that the target remains in the same location on the detector array. As the calibration pipeline algorithms can be computationally intensive on the lengthy exposures that are typical for TSOs, data files are segmented by the data management system to a maximum file size of $\sim$2 GB. The data products for each segment contain a dedicated time stamp extension (INT\_TIMES), which captures the start and end time of each integration, and which allows the calibration pipeline (and the user) to reconstruct the timing of the observation. The \texttt{jwst} calibration pipeline has dedicated procedures for the TSO modes to optimise the processing and calibration procedures for the highest precision (relative) spectro-photometric precision. Most notably, the TSO pipeline treats each integration as a separate observation ``unit'' for calibration purposes, rather than co-adding them to maximize signal to noise. The Stage 3 TSO pipeline merges the exposure segments and produces time series-specific output products, such as a white light curve for spectroscopy and a photometric time series for imaging TSO modes. 

\noindent Pre-launch estimates of the instruments' performance for high precision spectro-photometry were highly uncertain, for a number of reasons. Detecting molecular signatures in exoplanet atmospheres through transit spectroscopy requires precision down to a few 10 parts per million (ppm); this is a very challenging level of stability to reproduce reliably with ground-based testing equipment. In addition, some critical systematic effects, such as the telescope pointing stability, can only be characterised in the space environment. To address this lack of ground-testing experience and provide an early in-flight performance benchmark, TSOs were carried out with all 4 science instruments during the initial 6-month commissioning period. The data presented in this paper are the outcome of these commissioning observations with MIRI.

\subsection{Transiting exoplanet observations with MIRI}
JWST MIRI~\citep{rieke2015} is particularly useful for characterizing exoplanet atmospheres. Strong bands of H$_2$O, CH$_4$, O$_3$, and NH$_3$ are found in the the $5 - 11$ $\mu$m wavelength region, and all but H$_2$O are difficult to observe with HST or other facilities. Observations at these mid-IR wavelengths are also expected to be less obscured by clouds and hazes than near-IR spectra that often have very muted spectral features \citep[e.g.,][]{Kreidberg2014, Morley2015}. In addition, as the only JWST instrument with coverage beyond 5~\micron, MIRI provides a crucial extended baseline in wavelength to the near-IR instruments, which can help break model degeneracies~\citep[e.g.,][]{barstow2015}.  \added{The MIRI instrument is especially well suited to characterize the thermal emission of warm ($T_\mathrm{eq} \approx 700 K$) and cool ($T_\mathrm{eq} \approx 500 K$) transiting exoplanets, which will only be detectable at mid-IR wavelengths \citep{Greene2016, Beichman2018, rieke2015}.}

\noindent MIRI offers three observational modes for TSOs: imaging for photometry, and the Low and Medium Resolution Spectrometers for spectroscopy. The Low Resolution Spectrometer~\citep[LRS;][]{kendrew2015, kendrew2018} can be used both with a fixed slit and in slitless mode; the latter is specifically opimtized for TSOs, and, as of cycle 1, is MIRI's prime mode for transit spectroscopy. 

\subsection{Operations of MIRI's slitless Low Resolution Spectrometer}

MIRI's LRS mode shares its optical path and focal plane with the MIRI imager~\citep{bouchet2015}. The spectral dispersion element, a Ge/ZnS double prism, is mounted in the imager filter wheel, providing a continuous spectrum from approximately 5 to 12~\micron~ with R$\sim$100. The spectral resolving power varies from R$\sim$40 at $\lambda\sim$5~\micron~ to R$\sim$160 at $\lambda\sim$10~\micron. The dispersion profile displays a turnover shortward of 4~\micron, effectively folding the spectrum back onto itself. A small spectral leak is present in the transmission profile of the double prism, lending the 2D slitless spectral images a ``dotted i'' appearance. For the fixed slit operation, a filter is mounted on the slit mask to block radiation below 4.5~\micron, thus avoiding the spectral leak and fold over issues. The full design and operations of the LRS is described in~\citet{kendrew2015}. 

\noindent While the LRS slit mode reads out the full MIRI imager detector array, slitless mode uses a dedicated subarray (SLITLESSPRISM), which measures 416 rows by 72 columns. The readout time for a single frame of the SLITLESSPRISM subarray only takes 0.159 seconds, substantially shorter than the readout time for the full detector of 2.775 seconds. The slitless mode of the LRS supports only the FASTR1 readout pattern, in which N$_\mathrm{groups}$-1 non-destructive frames are executed followed by a read-reset at the end of an integration, yielding the total of N$_\mathrm{groups}$ reads. In multiple integration data, as the TSOs are, the read-reset is followed by an additional reset frame which takes an additional 0.159 seconds.

\noindent The slitless mode offers the following advantages for such observations over operation of the fixed-slit LRS mode: (i) first, the absence of a slit avoids pointing-induced flux losses; and (ii) the shorter subarray readout time increases the dynamic range and allows observations of brighter targets (by a factor of $\sim$17 compared with full array readout). The lack of a slit mask, conversely, allows more background to enter the aperture, and the overall sensitivity of the LRS in slitless mode is approximately an order of magnitude lower than when the slit is used. For transit spectroscopy, however, which is usually performed on bright exoplanet host stars, the gains in stability and dynamic range outweigh this loss of sensitivity.

\noindent Target acquisition (TA) is a mandatory part of LRS slitless observations, and this follows the same procedure as all TA sequences in MIRI. The TA target is first placed in a dedicated $48 \times 48$~pixel aperture ($\sim 5 \times 5$ arcsec) box, using one of 4 possible TA filters: F560W, F1000W, F1500W or the neutral density filter FND. An on-board centroiding algorithm determines the centre of mass of the target within the region of interest, and computes the required offset to the nominal pointing location. The telescope then executes a small angle manoeuvre (SAM) to place the target. Prior to selecting the double prism in the filter wheel (P750L), a TA verification image is taken, allowing the user to visualise where the target was placed before dispersion. 

\noindent The overall performance of MIRI's LRS mode will be described comprehensively in a future publication. In this work, we focus specifically on the data characteristics and performance of the mode for TSOs.

\section{Observations and basic calibration}\label{sec:observations}

As part of the MIRI commissioning program, we performed a spectroscopic time series observation of the transiting super-Earth \object{L 168-9 b} (also known as TOI-134) as part of the LRS photometric sensitivity and stability activity (PID 1033). 

The goals of the observation were to:

\begin{enumerate}
    \item test the Astronomer's Proposal Tool (APT) template for time series observations with the MIRI LRS mode, in particular with regards to the timing and exposure windows;
    \item test the observatory performance for acquisition and pointing stability, over characteristic timescales of primary transits or secondary eclipses;
    \item test and optimize the end-to-end performance of the \texttt{jwst} data calibration and processing pipeline; 
    \item investigate the impact of different calibration steps on the spectro-photometric precision and determine a ``recipe'' for optimal calibration to provide to the community; and
    \item provide a first in-flight estimate of the spectro-photometric noise floor of the MIRI slitless LRS.
\end{enumerate}

In this section we describe the target selection rationale, and address goals 1 through 3 of the above list. 

\subsection{Performance goals and target selection}

We set two main criteria against which to test the performance of MIRI LRS in slitless mode for transiting exoplanet observations. The first is for LRS data have less than 100 ppm systematic noise per R=50 bin in the spectrum of a bright transiting planet at 7.5 $\mu$m wavelength. We chose this wavelength because it is close to the middle of the high SNR region of the LRS range, and it is also very close to the strong 7.6 $\mu$m CH$_4$ band that will be observed in the spectra of cool transiting planets. The 100 ppm noise level limit is considerably higher than the broadband systematic noise measured in long-duration Spitzer IRAC 8~\micron~time series data \citep{Knutson2009}, but would still be useful for retrieving molecular abundances in a variety of transiting planets \citep[e.g.,][]{Greene2016}. The second goal was that the ``white'' light curve of the observations (i.e. the lightcurve integrated over the full LRS spectral range) would recover the transit depth of a known planet to within 100 ppm. Simultaneously meeting both goals would ensure that MIRI LRS produces data that was both sufficiently precise and accurate for scientific analyses of exoplanet atmospheres.

\noindent These performance goals led to our target selection criteria:
\begin{enumerate}
\item The host star is bright enough to have less than 70 ppm photon noise (1 $\sigma$) in its planet's transit spectrum binned to R=50 at $\lambda = 7.5~\mu$m. This would allow measurement of any instrument noise down to the $\sim$30 ppm level. The host star should also not saturate the detector in 5 or fewer FAST1R groups, the recommended minimum number for MIRI integrations.
\item The planet should be small enough in radius, have high enough surface gravity, and/or a low enough T$_{eq}$ so that it is expected to have a transmission signal amplitude comparable to the star's photon noise in the LRS bandpass. This requirement is intended to ensure that any features observed in the transmission spectrum are likely to be measurements of instrument noise and not spectral features that originate in the planet's atmosphere.
\item The system should have a well-known transit signal (known to  $1\sigma \simeq 100$ ppm or lower) that we will measure in the white light curve of the time series data.
\item The planet's transit duration should be T$_{14} \lesssim 2$ h with a period P $<$ a few days to minimize observing time and to enable flexible scheduling.
\item The system should be observable by JWST in May-June 2022 when the observations were likely to be scheduled.
\end{enumerate}
We found that \object{L~168–9~b} / \object{TOI~134~b} \citep{astudillo_defru2020} was the only exoplanet system that met all of these requirements. The planet orbits an M1V star at a distance of $\sim 25$ pc, with a period of 1.4 days. The planet's estimated mass is 4.6 $\pm$ 0.56 $M_E$. From the planet's discovery parameters, the uncertainty on the transit epoch at the likely time of observation was $\sim4$ hours. However, this improved to 11 minutes with new transit ephemerides computed from all TESS observations provided by B. Edwards (private communication). Theses new P and T0 ephemerides are similar to those recently published by \citet{patel2022}, and the two sets predicted transits times 9 minutes apart on in late May 2022 when the MIRI observations were scheduled.

\subsection{Observation details}

We used the JWST Exposure Time Calculator v1.6 \citep[ETC;][]{JWST_etc2016}\footnote{\url{https://jwst.etc.stsci.edu/}} to estimate the basic detector settings; in particular the number of groups per integration. This number sets the length of each integration and thus effectively the cadence of the time series, and determines the SNR that will be achieved in each integration. For L~168-9~b we find an optimal NGroups setting of 9, which will avoid any pixels reaching saturation level. The host star was modelled in the ETC using an M-dwarf Phoenix stellar model normalised to K magnitude of 7.1. 

L~168-9 was observed on 2022 May 29 UT in a time-series exposure of 9371 integrations lasting 4.14 h, starting approximately 2.8 h before the center of the L~168-9~b transit estimated to occur at 11:00 UT. This total time duration includes the out-of-transit time, $\sim$30 min of detector settling time, and additional time to accommodate scheduling flexibility. The APT file can be retrieved using PID 1033\footnote{\url{https://www.stsci.edu/jwst/science-execution/program-information.html?id=1033}}, and the data are publicly available in MAST under the same program identifier (Observation 5). The observation is divided into 5 segments. 

At the start of the visit, the target was placed at the nominal pointing position in the SLITLESSPRISM subarray with a target acquisition sequence, using the target itself for TA and the F1000W filter. The TA verification image showed that the target was placed at $dx, dy = (-0.14, -0.02)$ px from the expected nominal location, which is consistent with the accuracy seen in other early slitless LRS activities. The x-offset (the cross-dispersion axis) has no consequence for the data calibration as long as pointing \textit{stability} is good during the exposure (see Section~\ref{sec:pointing}). The y-offset can introduce a wavelength calibration error; however the placement accuracy achieved here, in combination with the excellent stability shown below, will not produce a significant error.  

 \subsection{Observatory pointing stability}\label{sec:pointing}
 
 To investigate the pointing stability of the spacecraft over the duration of the time series exposure, we retrieved engineering telemetry data for the Fine Guidance Sensor \citep[FGS;][]{doi:10.5589/q09-006} and the HGA. The FGS records high-precision positions and fluxes for the guide star at a 64~ms cadence; these measurement provide valuable data for decorrelating observed changes in the time series spectra. The HGA is permitted to move over the course of a lengthy time series observation, and the telemetry for these moves is equally valuable for diagnosing observed jumps. The positional FGS mnemonics are \textit{SA\_ZFGGSPOSX}, \textit{SA\_ZFGGSPOSY}, and the HGA move mnemonic is \textit{SA\_ZHGAUPST}; these are accessible via MAST.
 
\begin{figure*}
     \centering
     \includegraphics[width=\textwidth]{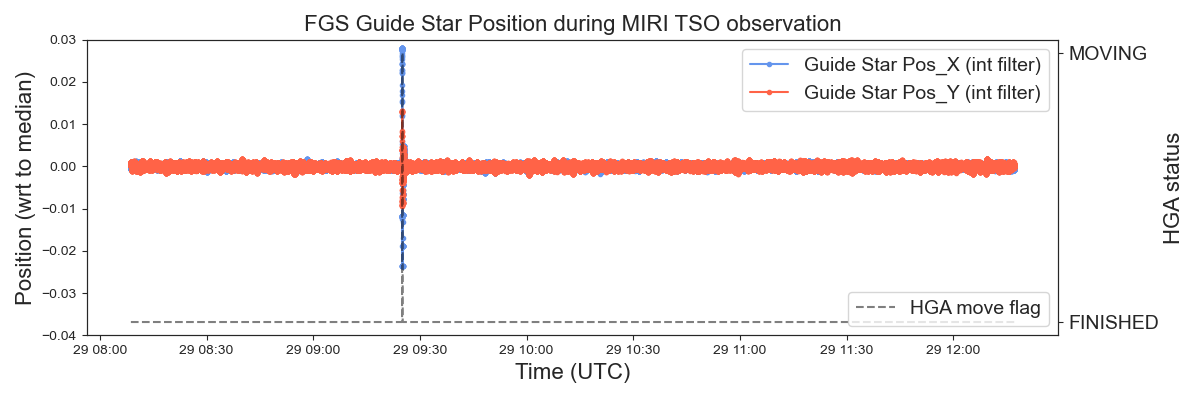}
     \caption{Engineering telemetry for FGS guide star position in X (blue) and Y (red), and for the High Gain Antenna (grey). The positional data are filtered to match the timescale of an integration in the MIRI time series observation. The HGA data show the observed position jump coincides very accurately with the brief HGA move.}
     \label{fig:engdb}
 \end{figure*}

 In Figure~\ref{fig:engdb} we show the time series of these mnemonics, filtered in time to match the cadence of our time series integrations. The trace shows exceptional pointing stability, with the standard error on the mean $6 \times 10^{-4}$ px in X and $4 \times 10^{-4}$ px in Y (relative to the median); the FGS pixel scaling is 0.069''/px. We observe just one event, where both positional measures are seen to deviate from the median. The HGA telemetry confirms that a move of the antenna took place at this time. Only a small number of data points are impacted by this event; these can be removed from the analysis with no impact on final measurement precision. 
 
\subsection{Basic calibration}
 
 Our processing started with the uncalibrated raw data \texttt{uncal} data products retrieved from Mikulski Archive for Space Telescopes (MAST). We used the Detector1 portion of the \texttt{jwst} calibration pipeline \citep[1.5.4.dev28+g254b2e6c;][]{jwst2022} for basic calibration, testing the impact of different steps on the resulting data quality and cosmetics; in this early stage of the mission, not all detector calibration files had been updated from their ground-based versions. For TSOs in particular, the most important calibrations are those that impact the stability of the time series. Effects that are highly stable between integrations do not necessarily degrade the precision of the time series. 
 
 The best calibration was obtained running the following sequence of steps: \texttt{dq init}, \texttt{saturation}, \texttt{reset}, \texttt{linearity}, \texttt{last frame}, \texttt{dark current}, \texttt{jump step} (with a modified threshold of 5.0) and \texttt{ramp fitting}. Skipped steps include some that do not apply to MIRI (e.g. \texttt{IPC}, \texttt{group scale}, \texttt{gain scale}); and others for which the correction has not yet been optimally derived for flight data (e.g. the \texttt{first frame}, and the reset switch charge decay (\texttt{RSCD}) steps). 
 
 The \texttt{jump} step is designed to detect cosmic ray hits, and uses a default detection threshold of 4. The algorithm computes differences between subsequent groups. With assumptions for Poisson and read noise, each 2-point difference is compared to the median, and outliers at $> x\sigma$, where $x$ is the detection threshold, are flagged. For our data we found the threshold of 4 flagged too many pixels erroneously; a threshold of 5 produced the best results. For very bright targets such as \object{L~168-9}, the steepness of the detector ramps possibly biases the algorithm towards over-detection of jumps; some trial and error is required to find the optimal rejection threshold for a particular dataset.  
 
 The \texttt{reset} step applies a correction for non-ideal behaviour of the instrument's Si:As detectors following a reset (which occurs prior to each integration). The first groups following a reset deviate from the expected linear accumulation of signal. This can take around 15 groups to settle in full array mode; longer in subarray. The reset step applies a correction to reduce this effect. 
 
 The MIRI detectors suffer from the classical non-linearity in the measured charge: there is a reduction in responsivity with increasing signal.  This non-linearity arises primarily due to the debiasing of the detectors as charge is accumulated on the amplifier integrating node.  For details on the theory of the non-linearity on the MIRI detectors, see~\citet{rieke2015_detectors}.
 The \texttt{linearity} step adjusts the integration ramps so the output of the adjusted DN is a linear function of the input signal.
 
 The \texttt{last frame} step makes no changes to the science or uncertainty extensions of the data, but flags the last group in each integration as ``DO NOT USE'' in the GROUPDQ array. The  MIRI detector is reset sequentially by row pairs. The last group of an integration ramp on a given pixel is influenced by signal coupled through the reset of the adjacent row pair. The result is that the odd and even rows both show anomalous offsets in the last read on an integration ramp. Including the last frame in the fit results gives an underestimation of the slope signal and imprints the odd-even row pattern. It's therefore prudent to exclude it from the fit. 

\begin{figure}
     \centering
     \includegraphics[width=0.45\textwidth,
     trim= 1.2in 0.0in 1.5in 0.0in, clip]{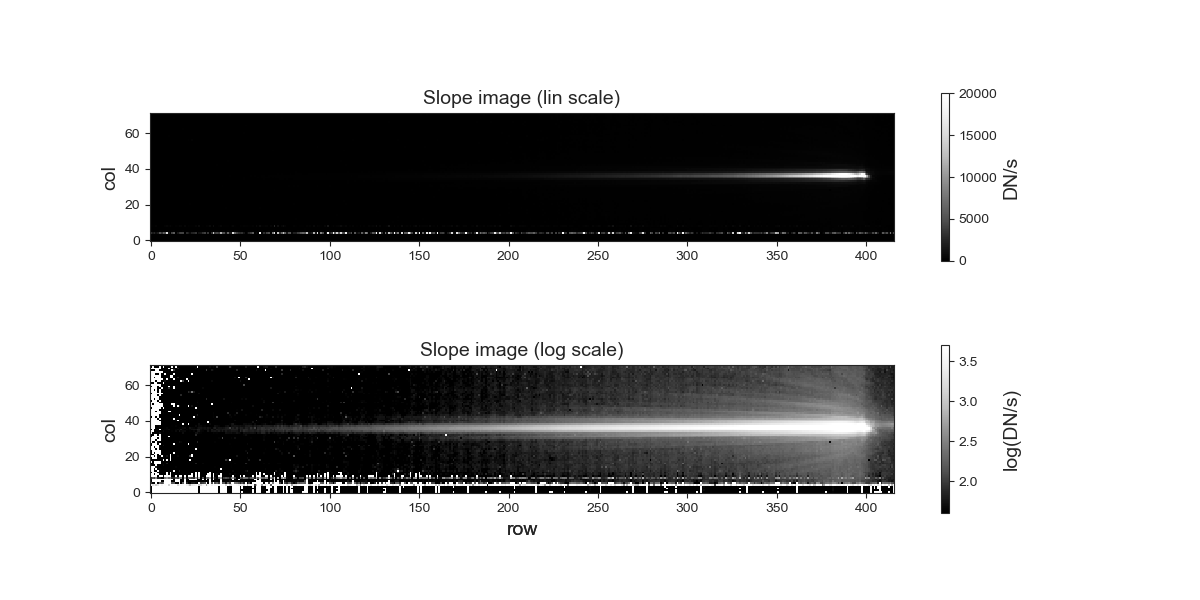}
     \hspace*{-2cm}
     \includegraphics[width=0.35\textwidth,
     trim=-0.1in 0.0in 2.5in 0.0in]{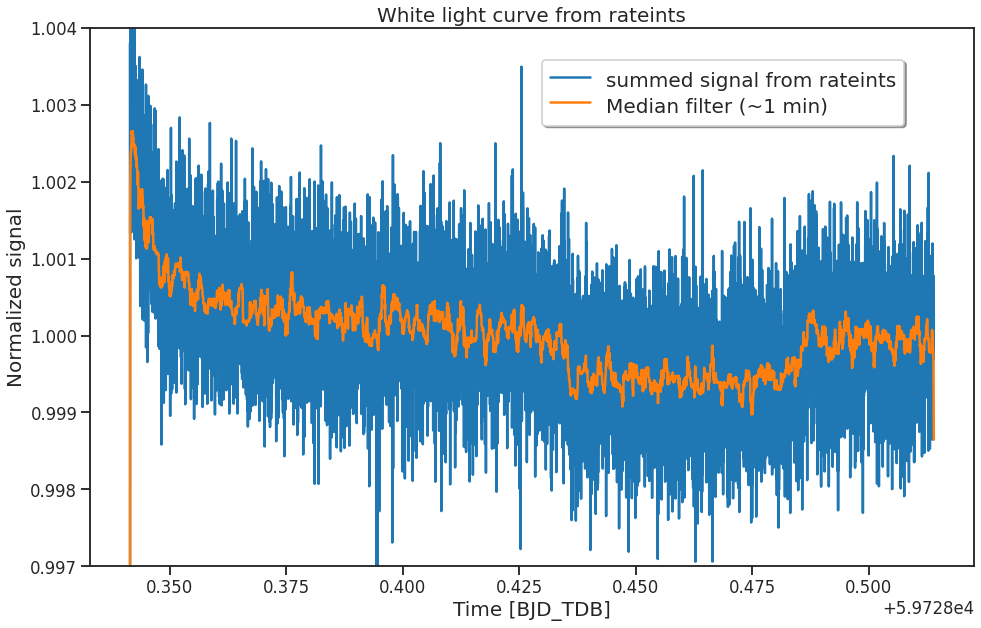}
     \caption{Level 2a calibrated spectral images and band averaged light curve of the \object{L~168-9} system. The images are rotated by 90 degrees for display purposes and the spectra are shown in the full SLITLESSPRISM subarray, which measures 416 rows $\times$ 72 columns on the detector. In this orientation, the wavelength increases from right to left. The images were generated from the seg-002 \texttt{rateints} file, which is one of 5 segments that make up the exposure. Both images show the same spectral image: (top) with a linear scaling; and (bottom) with a log scaling and limits adjusted to bring out the complex diffraction patterns and scattered light seen at the short wavelengths. The normalized lightcurve is obtained by summing the signal seen in the spectral images between detector rows 230 to 380 and detector columns 20 to 51 and dividing the total signal per integration by the averages signal in time. Even in this very basic calibrated data the planetary transit can be clearly observed.}
     \label{fig:rateints_ims}
 \end{figure}

 Both analysis methods presented in this paper use the same basic calibrations as are described in this section. Figure~\ref{fig:rateints_ims} shows an example of a Level 2a calibrated spectral image on the detector array. Note that in addition to the marked diffraction features seen in the image (highlighted particularly in the log scales image in the bottom panel of Figure~\ref{fig:rateints_ims}, we also see an additional scattered light component at the short-wavelength side of the trace. This is the manifestation in LRS of the so-called MIRI cruciform artifact, which is described in detail in~\citet{gaspar2021}, and discussed further in Section~\ref{sec:transit_noise}. At this stage of the pipeline calibration, the images are provided in units of DN/s. The lower panel of Figure~\ref{fig:rateints_ims} shows the white light curve based on a simple summation of the Level 2a calibrated spectral data.  The transit of L~168-9~b can be clearly seen, even without removing any of the systematic noise in this dataset, demonstrating the excellent photometric stability of the LRS. The largest response drifts are observed during the first $\approx 20$~minutes, at the level of about 0.3~\%. The rest of the data shows only a small, almost linear drift at the level of about 500~ppm.  
 
\section{Spectral extraction and time series analysis}\label{sec:analysis}

\subsection{\texttt{CASCADe} Analysis}

For the second stage of the data reduction pipeline (the \texttt{Spec2Pipeline}) we again largely use the default \texttt{jwst} calibration pipeline, with two step modifications. Our stage~2 processing starts with the  \textit{rateints} data product from the Detector~1 pipeline stage. For the steps of the Stage~2 \texttt{jwst} pipeline used in our data calibration we used the calibration and reference files corresponding to those available in the JWST Calibration Reference Data System (CRDS) with the pipeline mapping version 859. The first Stage 2 pipeline steps we applied were the \texttt{assign wcs} and \texttt{flat field} steps, assigning to each pixel in the spectral images a wavelength and RA and DEC coordinate, and correcting for differences in pixel response, respectively. We did not apply the \texttt{photom} step as absolute flux calibration is not needed for the determination of the relative transit depth. \added{Note that the \texttt{flat field} step could in principle also have been skipped for this particular data set without out loss of relative photometric precision due to the excellent pointing stability of JWST during the observation.}

\begin{figure}
\centering
\includegraphics[width=0.95\linewidth]{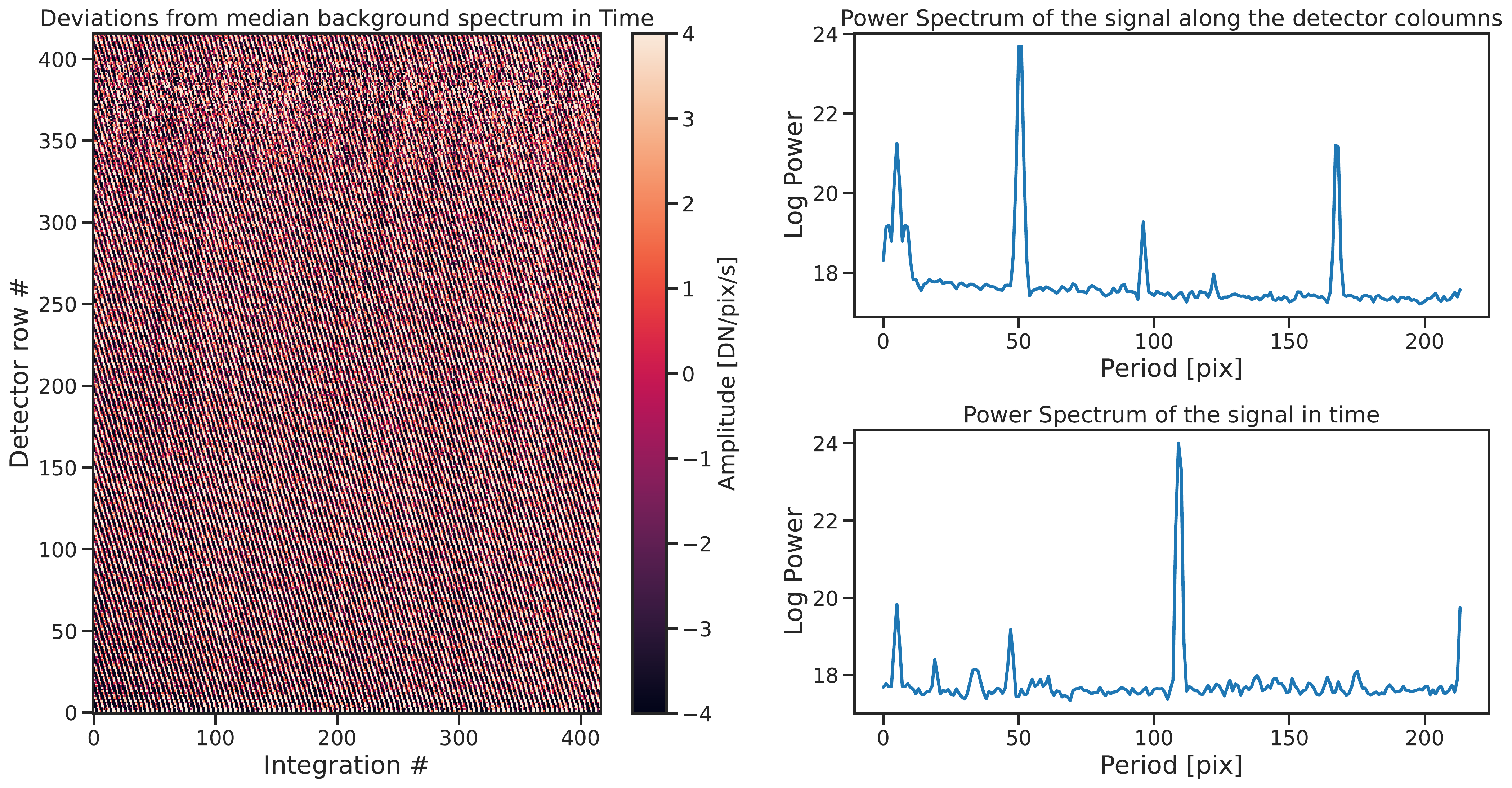}
\caption{Periodic correlated noise in the time series observations of \object{L~168-9~b}. The left panel shows the median subtracted background spectrum for the first 416 integrations. Plotted on the y-axis is the detector row number, i.e. in the dispersion direction of the spectrograph, the x-axis the integration number. The right panels show the power spectrum of the Fourier analysis of the left figure in the directions along the, respectively, y- and x-axis.
\label{fig:periodic_noise}}
\end{figure}

The current default pipeline does not yet contain a default background subtraction method for LRS slitless data, so we investigated during our analysis the optimal way of subtracting the background. We subtracted the infrared background emission in the spectral images for each integration separately. For this we determined a median background per integration by calculating the median in the cross-dispersion direction for detector columns 10 to 17 and 57 to 72. The median background spectrum is then subtracted from each detector column of the spectral images. Note that we do not use the detector columns 1 to 9, as the first 4 columns are reference pixels, and columns 5 to 9 exhibit an excess noise. A time dependent background correction , i.e. per integration, is essential as the time series data of \object{L~168-9~b} exhibits a periodic noise source. Figure~\ref{fig:periodic_noise} shows the normalized median background spectrum for the first 416 integrations together with a Fourier analysis, clearly showing the periodic modulations in this data set. Our current understanding is that the observed modulation of the detector signals arises from an as-yet unknown interaction between the detector control voltage generator circuits and the clocking waveform generator. Research is ongoing to eliminate or at least mitigate the noise in the raw data. Subtracting the background per integration effectively removes the observed periodic noise of Figure~\ref{fig:periodic_noise}, in a similar way as the $1/f$~noise observed in the timeseries observations with near-infrared instruments like NIRCAM \citep{2020AJ....160..231S}. 

After the background subtraction we performed an additional cosmic hit and bad pixel search and correction. For this we used the \texttt{CASCADe-filtering}\footnote{\url{https://pypi.org/project/CASCADe-filtering/}} package which is a sub package of the  Calibration of trAnsit Spectroscopy using CAusal Data (\texttt{CASCADe}\footnote{\url{https://pypi.org/project/CASCADe-spectroscopy/}}) data reduction package developed with the  \emph{Exoplanet Atmosphere New Emission Transmission Spectra Analysis} (\texttt{ExoplANETS-A}\footnote{\url{https://cordis.europa.eu/project/rcn/212911/factsheet/en}}) Horizon-2020 program. The applied filtering method has previously been used on HST transit spectroscopy data \citep{2021A&A...646A.168C}. We refer to this later paper for further details on the method. In brief, the applied filter
method assigns an optimal filter kernel to each detector pixel such that dispersion profile of the source is not broadened in the the spatial direction.
We flag all pixels deviating more than 4.5-$\sigma$ from the mean determined by the applied an-isotropic optimal filter profile. The flagged pixels are then replaced by an interpolated value using the same filter profile. Note that the filtering is applied on each spectral image in the timeseries separately, to preserve all temporal information and not flag and remove real time variable signals other than cosmic hits.

After cleaning the spectral images we determined the location of the spectral trace. For this we used the \texttt{CASCADe-jitter}\footnote{\url{https://pypi.org/project/CASCADe-jitter/}} package, also a sub package of the \texttt{CASCADe} transit spectroscopy package. This package uses an implementation of the Canny edge filter \citep{canny_filter} to create a binary image of the spectral images. The Jacobian and Hessian matrices needed for this method are calculated using a  
3x3 Scharr-Operator \citep{scharr2000}. The sub-pixel position of the spectral trace is determined by making a second order Taylor expansion of the Hessian matrix in the direction of the maximum value eigenvector (i.e. perpendicular to the direction of the spectral trace) to determine maximum of second order derivative, i.e. maximum of the spectral trace, for those pixels identified to be part of the spectral trace by the Canny edge filter. We then fitted a third
order polynomial to the derived sub pixel positions of the spectral trace. The time averaged polynomial coefficients of the spectral trace are $35.12$, $2.117 \times 10^{-3}$,  $8.657 \times 10^{-6}$ and $-1.890 \times 10^{-8}$ from zero to third order, respectively. We find an excellent agreement with the average trace position as a function of time and the FGS guide star positions as shown in Figure~\ref{fig:engdb}, and also showing no evidence of substantial telescope movement apart from the the brief period during the HGA movement.

Finally we extracted the 1D spectral timeseries data from the spectral images using the \texttt{extract1d} pipeline step with a custom parameters file. We apply the polynomial coefficients from the trace fit listed above to center the extraction aperture at the exact source position for all wavelengths. In the spectral extraction we used a constant width extraction aperture of 8 pixels.
\added{We empirically determined that for an constant width extraction aperture extraction width of 8 pixels gave the optimal noise performance and spectral stability. Future improvements on this could be a tapered extraction aperture to take into account the wavelength dependent PSF width, or a PSF weighted extraction such as discussed in section~3.2.}

\begin{figure*}
\includegraphics[width=0.95\linewidth]{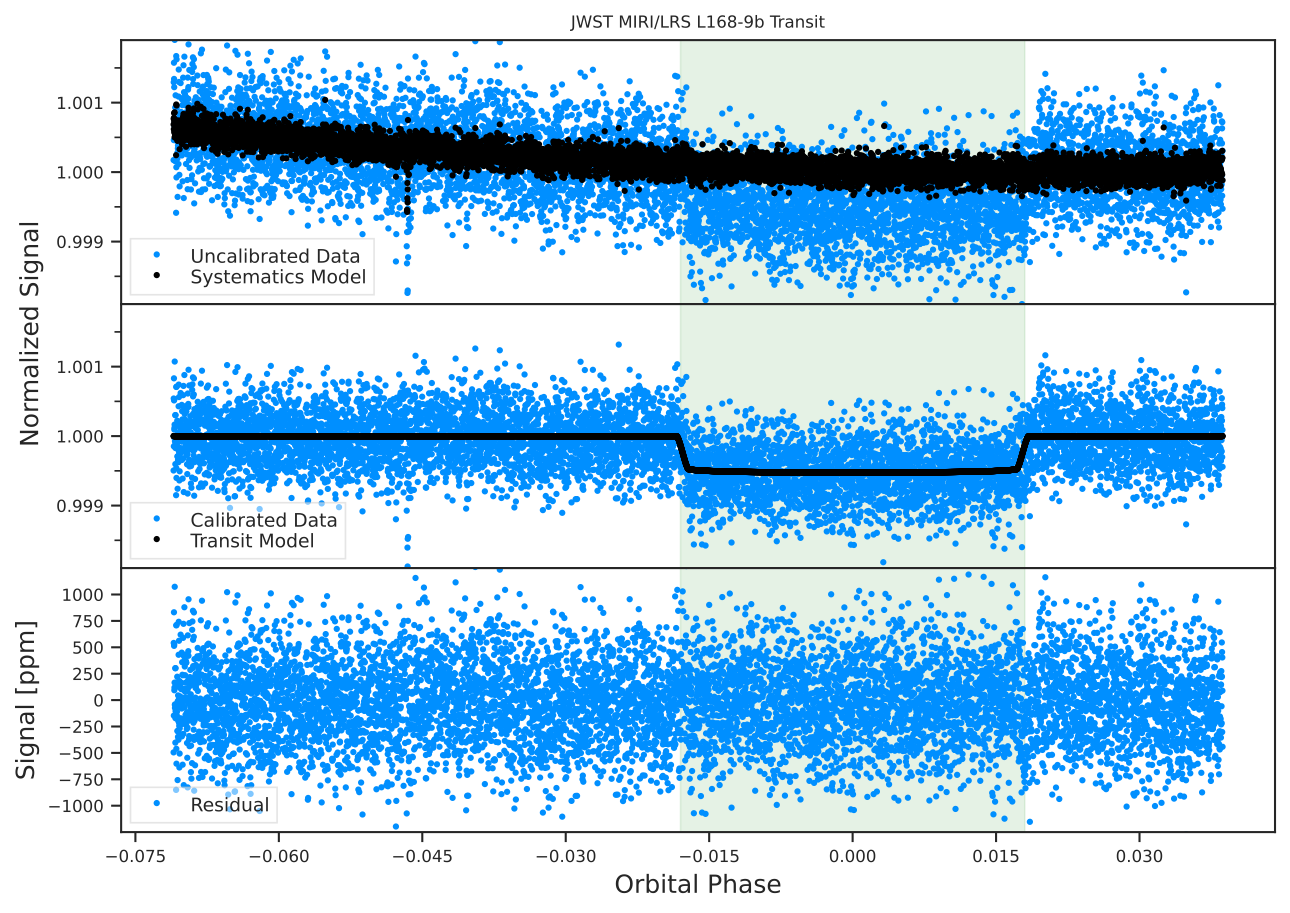}
\caption{\texttt{CASCADe}'s white-light (4.85--11.85~$\mu$m) light curve analysis.  The top panel shows the uncalibrated white light curve (blue data points) together with the fitted band averaged systematics model (black data points). The middle panel show the calibrated data together with the fitted light curve model. The lower panel shows the residuals from subtracting the lightcurve model from the calibrated time series data.  In the analysis the first 30 minutes of the observations have been omitted as they sow the strongest response drifts (which can be seen in Figure~\ref{fig:rateints_ims}).
\label{fig:fig_LC}}
\end{figure*}

The resulting band-averaged timeseries data is shown in the top panel of Figure~\ref{fig:fig_LC}.

To calibrate the extracted spectral timeseries data and to extract the transmission spectrum of L~168-9~b we used the \texttt{CASCADe} transit spectroscopy package.  This package makes use of the \emph{half-sibling-regression} methodology developed by \cite{Schoelkopf7391} using causal connections within the dataset to model both the transit signal and any systematics, and which has been successfully applied to transit observations from the Kepler mission \citep{Wang2016PASP} and field-stabilized imaging data \citep{Samland2020}. The \texttt{CASCADe} package was successfully applied to HST spectroscopic timeseries data \citep{2021A&A...646A.168C}. We refer to this latter paper for details on the half-sibling-regression method and its implementation onto the \texttt{CASCADe} package.

\begin{figure}
\centering
\includegraphics[width=0.9\linewidth]{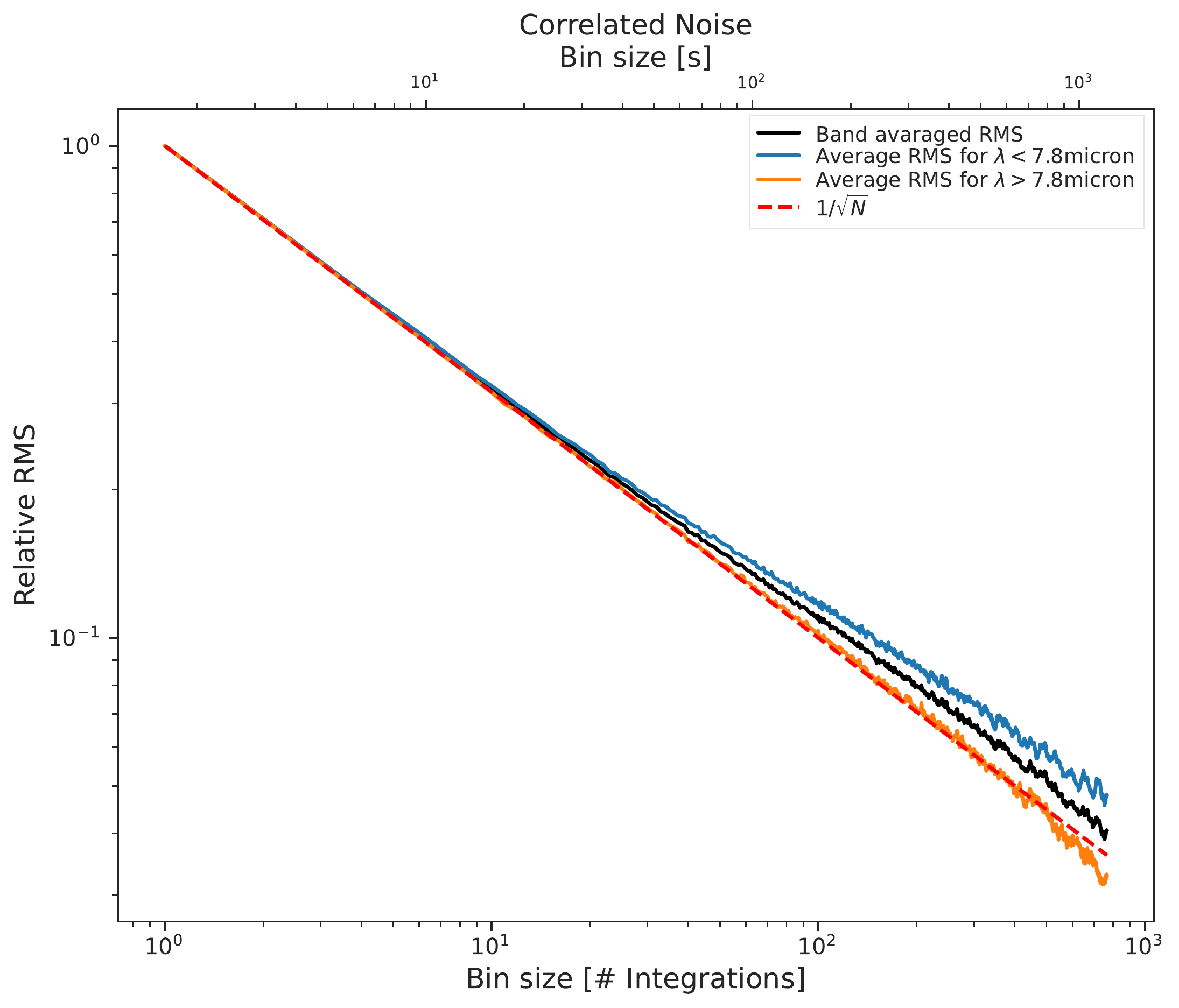}
\caption{The Allan variance plot \citep{Allan1966} for \texttt{CASCADe}'s light curve fits.
The black curve shows the relative RMS for the band averaged fit residual, and the red dashed line the expected $1/\sqrt{N}$ behaviour in case of uncorrelated noise. \added{ The absolute RMS for the band averaged residual, used to scale this curve is 331~ppm.} The orange line shows the Allan variance for wavelengths larger than 7.8~$\mu$m, while the blue line shows the Allan variance for wavelengths shorter than 7.8~$\mu$m. From this, it is clear that the shorter wavelength data still has a small systematic component.
\label{fig:cascade_allan}}
\end{figure}

Before fitting the spectral lightcurves, we rebinned the spectra to a uniform wavelength grid with a spectral resolution of 50 at 7.5~$\mu m$. The center values of the wavelength bins are listed in Table~\ref{tab:jwst_spectra}. Further, the spectral timeseries data shows a strong response drift during approximately the  first 30 minutes as can be seen in Figure~\ref{fig:rateints_ims}. In our analysis we, therefore, decided to skip the first 1019 integrations of the time series showing the strongest systematics. Our analysis of the remaining three and a half hours of the timeseries data only reveals a small systematic signal drift as shown in the top panel of Figure~\ref{fig:fig_LC}.

The lightcurve fitting with the \texttt{CASCADe} package currently only fits for the transit depth, all other parameters are kept fixed at an appropriate value. We assumed an eccentricity $e = 0$ and an inclination $i = 85.5^{\circ}$, values taken from the L~168-9~b discovery paper \citep{astudillo_defru2020}. 
We further used a mid transit time  $\mathrm{T}_0 = 2459728.9599$~d, corresponding to the mid-transit time seen in our data, and a relative semi-mayor axis $a/R_s = 7.3469$,  the later is within 1 sigma from previous reported values \citep{astudillo_defru2020,patel2022}.  Both values where estimated by requiring a continuous systematics model over the ingress and egress of the transit.
For the limb-darkening correction, we used the non-linear limb-darkening law proposed by \cite{2000A&A...363.1081C} of which the four wavelength dependent limb darkening coefficients were calculated using the \texttt{ExoTETHyS} package \citep{2020AJ....159...75M} and the  Atlas stellar model grid \citep{2000A&A...363.1081C}.  The lightcurve models we fitted to the data were calculated using the Batman package \citep{batman2015} using the above parameters as input. In our systematics modeling \cite[see][for details]{2021A&A...646A.168C} we used as an additional regressor a second order polynomial as a function of time and the time dependent trace position determined by the \texttt{CASCADe-jitter} package (see above). The errors on the transit depth fit and systematics model were estimated using a bootstrap sampling using 600 samples. 

Figure~\ref{fig:fig_LC} shows the fitted band averaged systematics model (top panel) and lightcurve model (middle panel). Note that the analysis method we applied fits both simultaneously and the calibrated lightcurve shown in the middle panel of Figure~\ref{fig:fig_LC} is not used for actual fitting but is derived from the combined fit to the uncalibrated data shown in the top panel. The bottom panel of Figure~\ref{fig:fig_LC} shows the residual after subtracting the fitted systematics and lightcurve model. 
The derived systematics show a slow downward trend, with a maximum range of about 500~ppm. The behaviour is not entirely linear in time and seems to reach a minimum and then remain constant after approximately 3 hours after the start of the observations. 
Figure~\ref{fig:systematics_per_wavelength_and_residuals} shows the systematics model and the residuals after subtracting the fitted systematics and lightcurve model for a selection of the wavelength channels corresponding to those listed in Table~\ref{tab:jwst_spectra}. For most wavelengths, the systematics behaviour is very similar, only for the longest wavelengths a difference can be observed i.e. a slight upward trend rather than a downward one. Shown in Figure~\ref{fig:cascade_allan} is the Allan variance plot \citep{Allan1966} for the residuals of the \texttt{CASCADe} lightcurve fit.
The Allan variance analysis for the band averaged residual only show a weak correlation, which is mainly caused by a correlated noise component at the shorter wavelengths. The longer wavelength show a behaviour consistent with the expected $1\sqrt{N}$ behaviour of uncorrelated noise.

The \texttt{CASCADe} transit spectrum is listed in Table~\ref{tab:jwst_spectra} and plotted in Figure~\ref{fig:fig_spectrum}. The band averaged transit depth is $524 \mathrm{ppm} \pm 15 \mathrm{ppm}$ 
or in relative planetary radii $\mathrm{R}_p/\mathrm{R}_s = 0.02288 \pm 0.00034$, which is consistent with the previous published values by \citep{astudillo_defru2020,patel2022} to within 1 sigma. \added{Note that our quoted errors are not adjusted for unmodelled correlated noise, as can seen in Figure~\ref{fig:cascade_allan} at a very low level for the shortest wavelengths.}

\begin{deluxetable*}{CC|CC|CC|CC}
\tabletypesize{\scriptsize}
\tablecaption{\texttt{CASCADe}'s transmission spectroscopy results. Each wavelength bin has a full-width of 0.149~$\mu$m.
\label{tab:jwst_spectra}}
\tablehead{\colhead{Wavelength} & \colhead{Transit Depth} &\colhead{Wavelength} & \colhead{Transit Depth} &\colhead{Wavelength} & \colhead{Transit Depth} &\colhead{Wavelength} & \colhead{Transit Depth}  \\
\colhead{($\mu$m)} & \colhead{(ppm)} &\colhead{($\mu$m)} & \colhead{(ppm)} &\colhead{($\mu$m)} & \colhead{(ppm)} &\colhead{($\mu$m)} & \colhead{(ppm)}  \
}
\startdata
    4.86 &  479 \pm  96 & 6.65 &  451 \pm  50 & 8.43 &  502 \pm  47 & 10.22 &  497 \pm   81 \\
    5.01 &  341 \pm  87 & 6.80 &  503 \pm  46 & 8.58 &  552 \pm  56 & 10.37 &  506 \pm   96 \\
    5.16 &  599 \pm  77 & 6.95 &  440 \pm  52 & 8.73 &  517 \pm  56 & 10.52 &  593 \pm   99 \\
    5.31 &  598 \pm  75 & 7.09 &  499 \pm  45 & 8.88 &  595 \pm  68 & 10.67 &  577 \pm  100 \\
    5.46 &  503 \pm  68 & 7.24 &  502 \pm  42 & 9.03 &  551 \pm  52 & 10.82 &  565 \pm  116 \\ 
    5.61 &  586 \pm  60 & 7.39 &  638 \pm  50 & 9.18 &  505 \pm  56 & 10.96 &  616 \pm  134 \\ 
    5.75 &  463 \pm  52 & 7.54 &  512 \pm  43 & 9.33 &  505 \pm  61 & 11.11 &  522 \pm  159 \\
    5.90 &  465 \pm  55 & 7.69 &  492 \pm  45 & 9.48 &  489 \pm  58 & 11.26 &  615 \pm  156 \\
    6.05 &  544 \pm  47 & 7.84 &  613 \pm  49 & 9.62 &  531 \pm  61 & 11.41 &  566 \pm  172 \\
    6.20 &  540 \pm  43 & 7.99 &  451 \pm  48 & 9.77 &  527 \pm  67 & 11.56 &  656 \pm  195 \\
    6.35 &  527 \pm  47 & 8.14 &  547 \pm  51 & 9.92 &  446 \pm  67 & 11.71 &  683 \pm  213 \\ 
    6.50 &  430 \pm  50 & 8.29 &  611 \pm  55 & 10.07 & 523 \pm  72 & 11.86 &  595 \pm  264 
\enddata
\end{deluxetable*}

\subsection{\texttt{Eureka!} Analysis}

We performed another independent reduction and analysis of the observations using version 0.5 of the \texttt{Eureka!} pipeline\footnote{\url{https://github.com/kevin218/Eureka/releases/tag/v0.5}} \citep{eureka2022}. The \texttt{Eureka!} analyses started with the same Stage 1 outputs as the \texttt{CASCADe} analysis. The ``Eureka!~Control Files" (ECFs) and ``Eureka!~Parameter Files" (EPFs) used in these analyses are available for download\footnote{{\color{red}Zenodo link to be created on final submission}} and are summarized below.

In Stage 2, version 11.16.5 of the \texttt{crds} package \citep{crds2022} with CRDS context 0928 and version 1.6.0 of the \texttt{jwst} pipeline were used. No changes were implemented relevant for MIRI LRS between versions 1.5.3 (used for the \texttt{CASCADe} analysis) and 1.6.0. The Stage 2 pipeline was run using default settings with the exception that the \texttt{photom} step was skipped as it is expected to degrade time-series observations. In Stage 3, a y-window of detector rows (140, 393) and x-window of detector columns (13, 64) were used to crop out noisy regions of the detector; the frame was rotated by 90 degrees to have wavelength increasing to the right; Gaussian centroiding was used; a source aperture half-width of 4 and background exclusion region half-widths of 8--16 were considered while a value of 10 was ultimately chosen as it provided spectral light curves with the lowest point-to-point scatter on average; a constant background value was fit and subtracted from each column with a 5 sigma clipping of bad pixels; and optimal extraction was performed using the median frame and using 10 sigma clipping while performing the optimal extraction. A bug was present in the \texttt{jwst} Stage 2 pipeline at the time of analysis which left the wavelength array in the Stage 2 \textit{calints} files unpopulated; as a result, wavelengths were computed using STScI provided code temporarily implemented\footnote{\url{https://github.com/kevin218/Eureka/blob/v0.5/src/eureka/S3_data_reduction/miri.py\#L184}} into \texttt{Eureka!}'s Stage 3.

In Stage 4, the spectra was binned into 48 spectral channels with equal widths of 0.149~$\mu$m spanning 4.86--11.86~$\mu$m as well as a white-light channel. The spatial centroid, spatial PSF-width, and extracted flux all showed a brief ($\sim$20 integration long) anomaly at the same time (near \mbox{BJD\textunderscore TDB} 2459728.39 or integration $\sim$2900) which was caused by the HGA move shown in Figure \ref{fig:engdb}; afterwards, all three parameters approximately settled back to their original values. To remove the impact of this event, we iteratively sigma clipped the light curves at 5 sigma using a box-car filter 500 integrations wide with a maximum of 10 iterations.

In Stage 5, \texttt{dynesty} \citep{dynesty_v1_2_3} was used to fit the observations using a \texttt{batman} transit model \citep{batman2015}, a double exponential ramp systematic model, a linear polynomial model to fit for the overall flux level and any linear slope in time, and a multiplier to the expected \added{white} noise level to account for any noise above the photon-limit as well as an incorrect value for the gain. \added{No additional error inflation was performed to account for unmodelled correlated noise.} First, we performed a fit to the white-light light curve. We fixed the orbital period and placed a Gaussian prior on the linear ephemeris, orbital inclination, and semi-major axis (in units of a/R$_{*}$) based on the values of \citet{patel2022}, and we assumed an eccentricity of zero; our astrophysical priors are summarized in Table \ref{tab:eureka_orbit}. Limb-darkening was fitted using the reparameterized quadratic limb-darkening method from \citet{kipping2013}. Two exponentially decaying ramps were fit to the data to remove the idle-recovery behaviour exhibited by the MIRI detector; to avoid degeneracies between the two exponential ramps, we used loose Gaussian priors based on an initial \texttt{scipy.optimize.minimize} fit to the white-light light curve. The median orbital period, linear ephemeris, orbital inclination, semi-major axis, and limb-darkening coefficients from the \texttt{dynesty} fit to the white-light observations were then set as fixed values in the spectroscopic fits to avoid variations in these wavelength-independent values causing additional noise in the final transmission spectrum. We also fixed the exponential ramp timescales to those fitted to the white-light fit as there was little evidence for wavelength-dependent timescales, although there is significant variation in the amplitudes of the ramps with wavelength. All other parameters were allowed to vary as a function of wavelength. All of our \texttt{dynesty} fits used 121 live points, ``multi'' bounds, ``auto'' sampling which resulted in random walk sampling for the white-light fit and uniform sampling for the spectroscopic fits, and were run until a dlogz value of 0.1 was achieved.

\begin{figure*}
\centering
\includegraphics[width=0.90\linewidth, trim=0 0 0 23, clip]{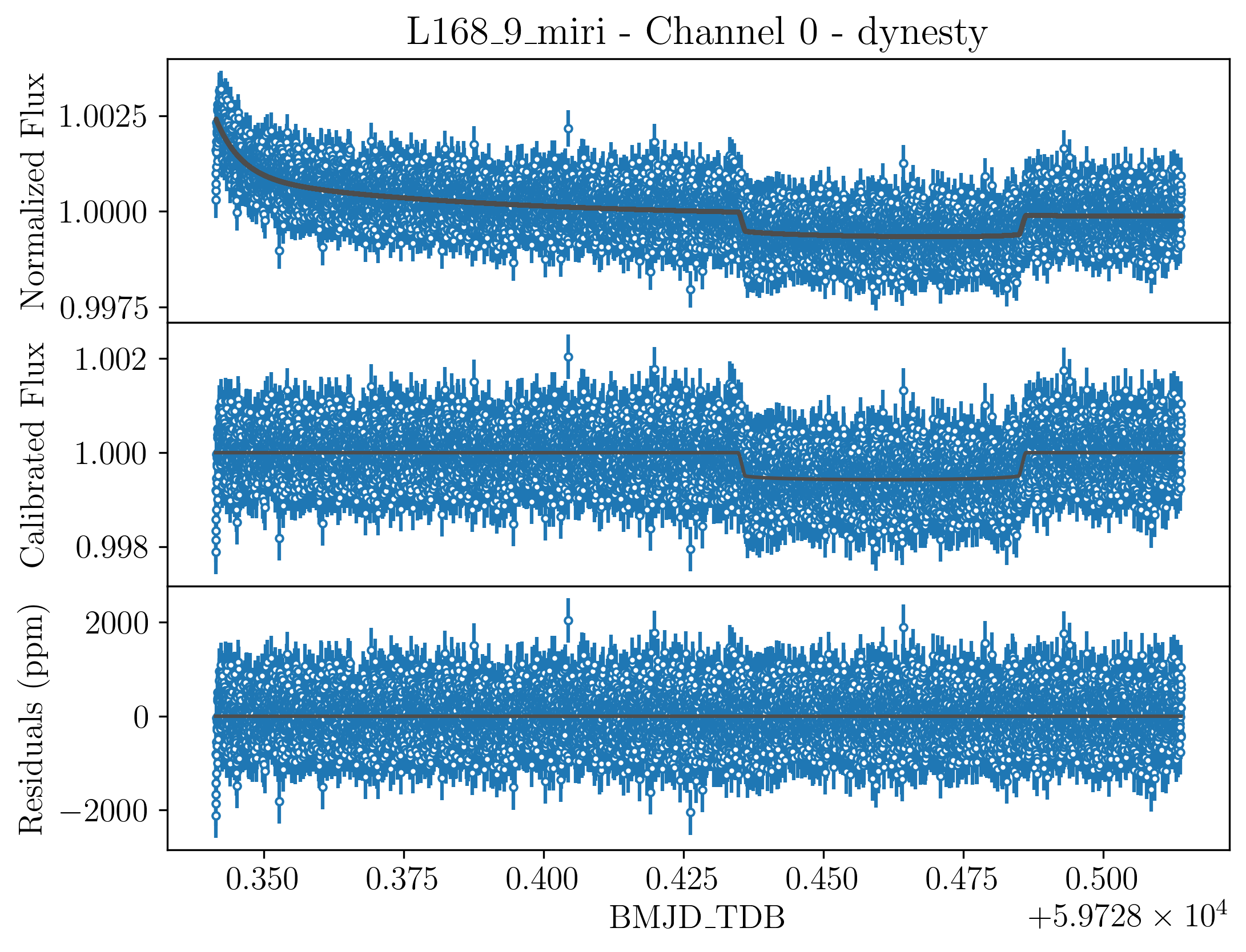}
\caption{\texttt{Eureka!}'s white-light (4.86--11.86~$\mu$m) light curve fit. Compared to Figure \ref{fig:fig_LC}, these data show a strong hook at the start of the observations because these initial integrations were not removed from the \texttt{Eureka!} analysis since they were well modelled using the double-exponential model.
\label{fig:eureka_lightcurve}}
\end{figure*}

\begin{figure}
\centering
\includegraphics[width=0.9\linewidth]{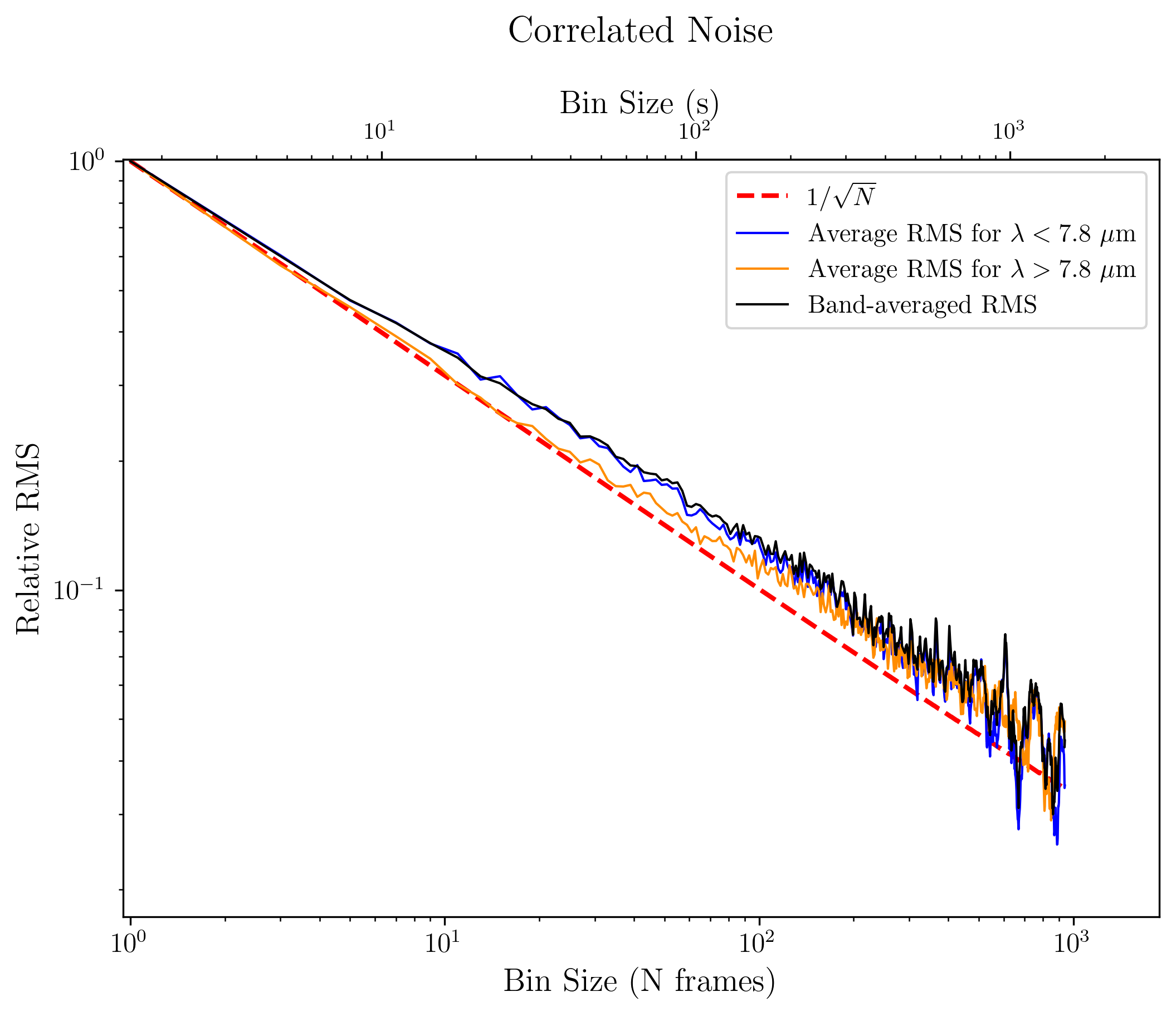}
\caption{The same plot as Figure \ref{fig:cascade_allan} but for \texttt{Eureka!}'s light curve fits. \texttt{Eureka!}'s shorter wavelength data also still has a small red noise component.
\label{fig:eureka_allan}}
\end{figure}

\texttt{Eureka!}'s fit to the white-light light curve is shown in Figure \ref{fig:eureka_lightcurve} with the Allan variance plot \citep{Allan1966} shown in Figure \ref{fig:eureka_allan}. Table \ref{tab:eureka_orbit} provides the following updated astrophysical parameters from the \texttt{Eureka!} analysis of the white light curve: transit ephemeris ($t_0$), inclination, $a/R_{\rm *}$, $R_{\rm p}/R_{\rm *}$, and limb darkening parameters. We find a transit depth of $547 \pm 13$ ppm for \texttt{Eureka!}'s white light curve, which is consistent within 1$\sigma$ of the value reported by \citet{patel2022} ($543 \pm 33$ ppm). \texttt{Eureka!}'s fitted transmission spectrum is tabulated in Table \ref{tab:eureka_spectrum} and plotted in Figure \ref{fig:fig_spectrum}. The average transit depth from these individual fits is $524 \pm 12$ ppm (where 12 ppm is the uncertainty on the mean) which is still very consistent with the value reported by \citet{patel2022}, and in excellent agreement with the \texttt{CASCADe} results.

\begin{deluxetable*}{C|CCCCCCC}
\tabletypesize{\scriptsize}
\tablecaption{\texttt{Eureka!}'s priors and fitted values for the white-light astrophysical parameters, where $u_1$ and $u_2$ are the limb-darkening parameters of the re-parameterized quadratic limb-darkening method from \citet{kipping2013}. A uniform prior is shown with $\mathcal{U}\rm(lower, upper)$, and a normal prior is shown with $\mathcal{N}(\mu,\sigma)$. Priors were based on the values of \citet{patel2022}.
\label{tab:eureka_orbit}}
\tablehead{ & \colhead{Period (days)} & \colhead{$t_{\rm 0}$ (BJD\textunderscore TDB)} &\colhead{inc ($^{\circ}$)} & \colhead{$a/R_{\rm *}$} & \colhead{$R_{\rm p}/R_{\rm *}$} &\colhead{$u_1$} & \colhead{$u_2$}
}
\startdata
\mathrm{Prior} & 1.4015272 & $\mathcal{N}(2459082.85700, 0.00040)$ & $\mathcal{N}(87.57, 1.7)$ & $\mathcal{N}(8.730, 1.125)$ & $\mathcal{N}(0.02, 0.005)$ & $\mathcal{U}(0,1)$ & $\mathcal{U}(0,1)$ \\
\mathrm{Fitted} & - & $2459082.856432^{+0.000089}_{-0.000084}$ & $88.3^{+1.0}_{-0.9}$ & $8.61^{+0.25}_{-0.37}$ & $0.02339^{+0.00028}_{-0.00027}$ & $0.034^{+0.038}_{-0.021}$ & $0.38^{+0.37}_{-0.24}$ 
\enddata
\end{deluxetable*}

\begin{deluxetable*}{CC|CC|CC|CC}
\tablecaption{\texttt{Eureka!}'s transmission spectroscopy results. Each wavelength bin has a full-width of 0.149~$\mu$m.
\label{tab:eureka_spectrum}}
\tablehead{\colhead{Wavelength} & \colhead{Transit Depth} &\colhead{Wavelength} & \colhead{Transit Depth} &\colhead{Wavelength} & \colhead{Transit Depth} &\colhead{Wavelength} & \colhead{Transit Depth}  \\
\colhead{($\mu$m)} & \colhead{(ppm)} &\colhead{($\mu$m)} & \colhead{(ppm)} &\colhead{($\mu$m)} & \colhead{(ppm)} &\colhead{($\mu$m)} & \colhead{(ppm)}
}
\startdata
4.86 & 560$^{+85}_{-83}$ & 6.65 & 695$^{+56}_{-61}$ & 8.43 & 441$^{+54}_{-53}$ & 10.22 & 670$^{+75}_{-90}$\\
5.01 & 364$^{+75}_{-68}$ & 6.80 & 516$^{+47}_{-45}$ & 8.58 & 565$^{+66}_{-60}$ & 10.37 & 562$^{+89}_{-94}$\\
5.16 & 557$^{+80}_{-78}$ & 6.95 & 399$^{+56}_{-54}$ & 8.73 & 521$^{+61}_{-65}$ & 10.52 & 524$^{+101}_{-89}$\\
5.31 & 565$^{+83}_{-77}$ & 7.09 & 536$^{+45}_{-52}$ & 8.88 & 658$^{+57}_{-55}$ & 10.67 & 529$^{+97}_{-89}$\\
5.46 & 490$^{+85}_{-82}$ & 7.24 & 451$^{+52}_{-50}$ & 9.03 & 535$^{+61}_{-55}$ & 10.82 & 480$^{+119}_{-106}$\\
5.61 & 657$^{+60}_{-55}$ & 7.39 & 683$^{+49}_{-53}$ & 9.18 & 548$^{+64}_{-69}$ & 10.96 & 442$^{+115}_{-109}$\\
5.75 & 410$^{+64}_{-59}$ & 7.54 & 595$^{+52}_{-54}$ & 9.33 & 477$^{+63}_{-55}$ & 11.11 & 642$^{+132}_{-121}$\\
5.90 & 456$^{+49}_{-55}$ & 7.69 & 433$^{+49}_{-47}$ & 9.48 & 481$^{+69}_{-73}$ & 11.26 & 410$^{+121}_{-115}$\\
6.05 & 538$^{+54}_{-52}$ & 7.84 & 620$^{+53}_{-54}$ & 9.62 & 618$^{+65}_{-64}$ & 11.41 & 382$^{+119}_{-127}$\\
6.20 & 519$^{+55}_{-52}$ & 7.99 & 507$^{+50}_{-45}$ & 9.77 & 564$^{+71}_{-79}$ & 11.56 & 433$^{+136}_{-135}$\\
6.35 & 559$^{+54}_{-53}$ & 8.14 & 547$^{+51}_{-53}$ & 9.92 & 478$^{+72}_{-70}$ & 11.71 & 449$^{+161}_{-144}$\\
6.50 & 463$^{+47}_{-48}$ & 8.29 & 572$^{+56}_{-55}$ & 10.07 & 446$^{+70}_{-72}$ & 11.86 & 623$^{+173}_{-172}$\\
\enddata
\end{deluxetable*}


\section{Discussion and performance summary }\label{sec:performance}

\subsection{White light curve}

The initial signal amplitude in these L 168-9 observations was approximately 0.25\% higher than the mean post-transit signal, and the excess signal decayed by a factor of $e$ within about 20 minutes. We find that the detector's idle-recovery behaviour in the white-light fit is well modelled by the double exponential ramp model used by the \texttt{Eureka!} analysis with one timescale of 68 minutes and another, shorter timescale of 6.5 minutes. The signal decay seen here should be typical for future MIRI LRS transit observations that also do not include significant latent images from previous exposures. However, the flux of the source and timing of the observation will have a strong influence on the shape and timescales of the settling observed. Indeed, the derived systematics model per spectral spectral channel from the \texttt{CASCADe} analysis as shown in Figure~\ref{fig:systematics_per_wavelength_and_residuals} indicates that the idle-recovery behaviour is different for the longest wavelength channels, seeing the lowest illumination levels in both background as well as L~168-9 itself.

The MIRI imaging / LRS detector was left in the low-background F560W filter and was not exposed to bright objects while being read out in an idling clock pattern for $\sim$8 hours before this observation. This ensures that latent images from previous exposures were not a significant component of the L~168-9~b time series observation. The first step of the observation is to move the filter to the F1000W filter for the target acquisition. The filter remained in F1000W for approximately 20 minutes allowing traps to begin to fill from this relatively high background filter. The double prism assembly (P750L) was rotated into the optical path about 75~s before the time series exposure started. The spectrum of L~168-9 was dispersed onto the detector during this time, allowing the detector signal levels to begin to stabilize to a new equilibrium while still being read out in an idling clock pattern. The detector readout pattern changed from idling resets to the nine-group FAST1R integration configuration at the start of the time-series exposure. The detector equilibrium signal level then changed to that of the nine-group readout, lower than the equilibrium level during idle reset mode. This can be explained by the increase in number of trapped charges during the integrations as we have reduced the number of destructive resets from idling to integrating. This is thought to be the cause of the downward or negative settling we see at the start of the exposure in the white-light timeseries. Such negative settling is consistent in amplitude, shape, and duration with response drift when switching from idling mode, on source, in pre-flight tests at NASA JPL of nearly identical detectors fabricated during the same production runs as the MIRI imager / LRS flight detector.

\added{Note that definite numbers of settling times and amplitudes, and recommendations to mitigate their effects, can only be made after sufficient MIRI LRS transit observations will have been made for a range of target brightness and observation duration. }

\subsection{Transit spectrum and Noise performance}\label{sec:transit_noise}
\begin{figure}
\includegraphics[width=0.95\linewidth]{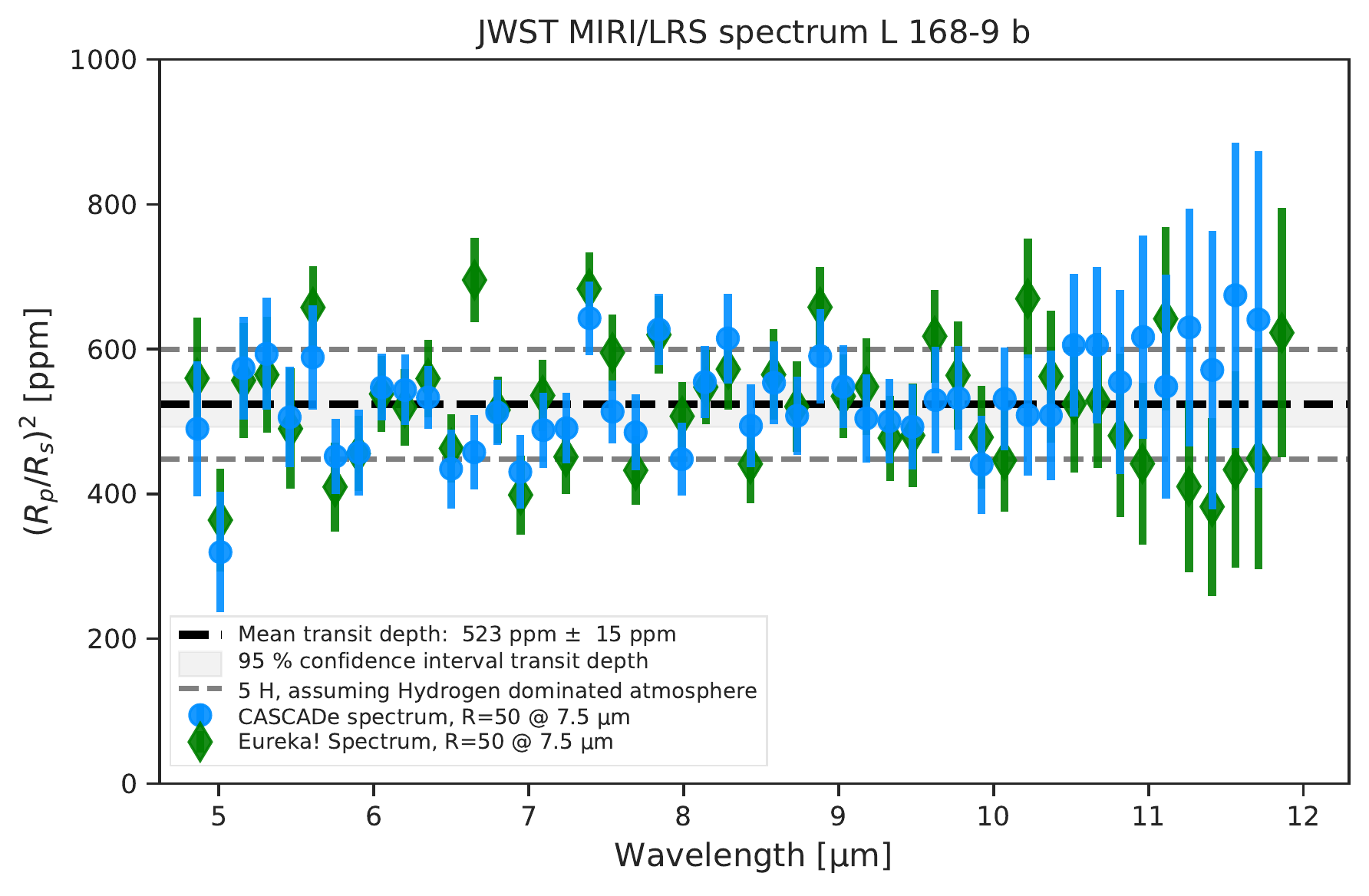}
\caption{The transit spectrum of L~168-9~b. The blue dots show the results from the analysis using the \texttt{CASCADe} code, while the green diamonds show the results from the analysis with \texttt{Eureka!}. The thick dashed line indicates the band averaged transit depth from the \texttt{CASCADe} analysis and the shaded area the 95$\%$ confidence interval. The thin dashed lines indicated five scale heights in the planetary atmosphere assuming a mean molecular weight of 2.4.  
\label{fig:fig_spectrum}}
\end{figure}

\begin{figure}
\includegraphics[width=0.95\linewidth]{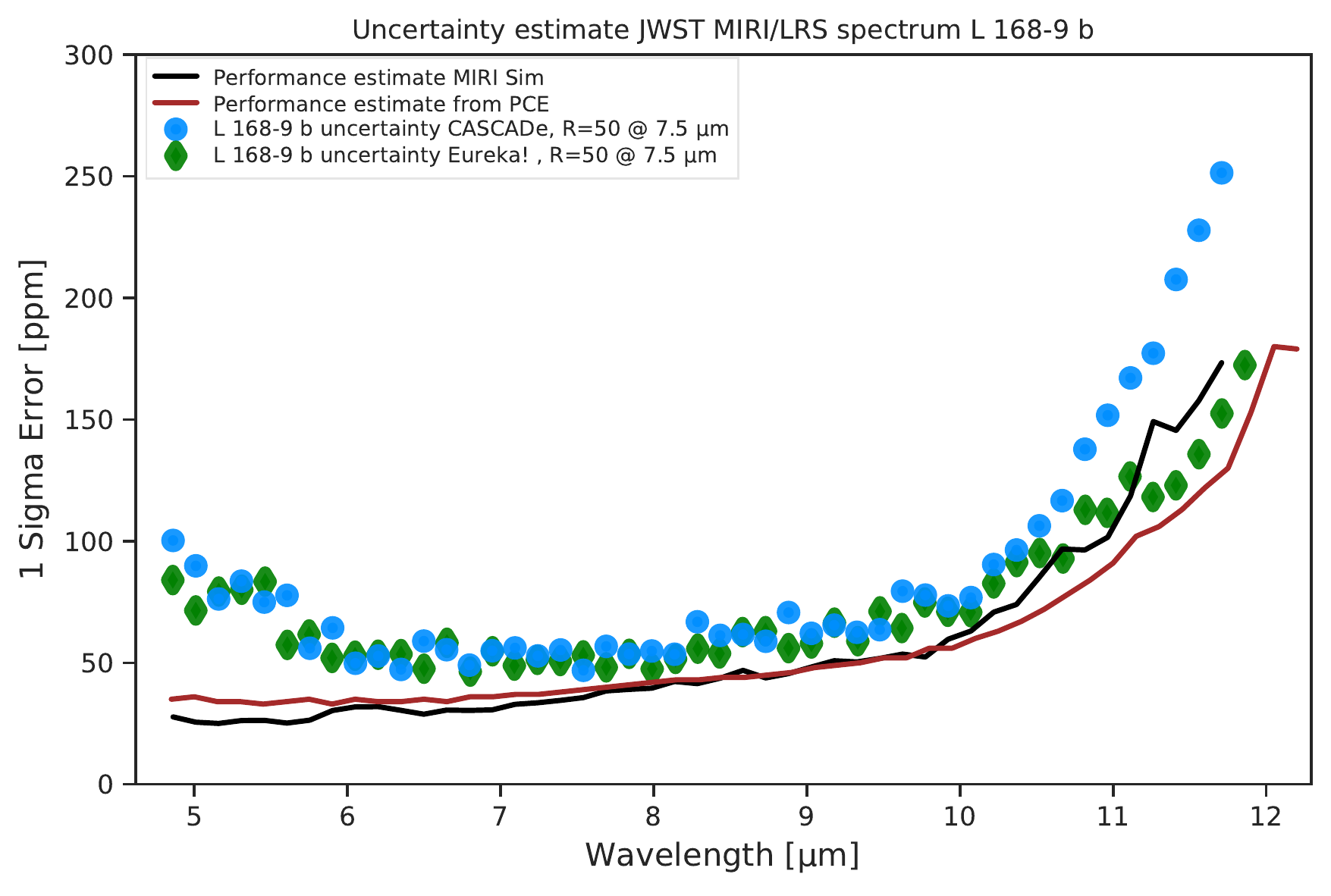}
\caption{\added{Noise estimates of the transit spectrum of L~168-9~b. The blue dots show \texttt{CASCADe}'s noise estimate while the green diamonds show \texttt{Eureka!}'s 1~$\sigma$ uncertainty estimate. The red curve shows the expected noise limit for this observations using the in-flight photon conversion efficiency \citep[see also][for the pre-flight PCE curve and derivation]{kendrew2015, Glasse2015}. The black curve shows the estimated noise limit using MIRISim \citep{mirisim} and the in-flight determined \texttt{PHOTOM} and wavelength calibration files (see also section \ref{sec:analysis}}).
\label{fig:fig_error_analysis}}
\end{figure}

The transmission spectra produced by \texttt{CASCADe} and \texttt{Eureka!} are shown in Figure \ref{fig:fig_spectrum}. They are generally quite similar and consistent with flat spectra within their estimated noise values. The uncertainties of each wavelength point (binned to 0.149 $\mu$m) are also similar for the \texttt{CASCADe} and \texttt{Eureka!} analyses. This is shown in Figure \ref{fig:fig_error_analysis} along with two independent predicted (simulated) noise estimates. One of these estimates was computed from simulated observations of the L~168-9 system with MIRISim \citep{mirisim}, using the in-orbit determined \texttt{PHOTOM} calibration file and PSF model. Based on these simulations, an error on the individual spectra in the spectral time series was estimated, which we used to simulate the light curves per spectral channel assuming a constant transit depth of 530~ppm. The simulated light curves were then fitted using the \texttt{CASCDe} package to estimate the error of the simulated transit spectrum. The other estimate (red curve in Figure \ref{fig:fig_error_analysis}) was computed from the estimated noise sources in the observation including source photon noise, background photon noise, and random detector noise sources within an 8 (spatial) $\times$ 6 (spectral) pixel extraction aperture ($R \sim$ 50). \added{This estimate assumed a background of 500 electrons s$^{-1}$ from version 2.0 of the \href{https://jwst.etc.stsci.edu/}{JWST exposure time calculator} (ETC) and the LRS in-flight photon conversion efficiency (PCE; also known as throughput) curve from that same ETC version.
The small differences between the two independent estimates of the noise floor are most likely due to small differences in assumed background levels and psf shapes in combination with the assumptions made for the spectral extraction.}

Figure \ref{fig:fig_error_analysis} shows that the measured noise is at, or slightly higher, than the simulated noise estimates over wavelengths $6.8 \lesssim \lambda \lesssim 11 \mu$m, but remains within approximately 20\% from these estimates. This suggests that any systematic noise in the observation is at most 30 -- 45 ppm at these wavelengths with the adopted bin width of 0.149 $\mu$m.

Compared to simulations, the observed spectrum has more noise than expected in the at wavelengths $\lambda \lesssim 7 \mu$m (see Figure \ref{fig:fig_error_analysis}). We have not positively identified the source of this excess noise, but we note that it corresponds to wavelengths that show excess scattering along the rows and columns of the detector array in Figure~\ref{fig:rateints_ims}. Radiation passes all the way through the detector substrate at these wavelengths and encounters multiple passes through its IR-sensitive layer and buried contact in addition to scattering along rows and columns between the Si detector substrate and its integrated readout (ROIC) \citep{rieke2015_detectors, gaspar2021}. We speculate that these multiple passes introduce excess noise, and the light also scatters into a cross-like image structure with some flux outside of our spectral extraction aperture. A possible other source for the observed excess noise could be the Reset Switch Charge Decay (RSCD) effect (see the MIRI technical note MIRI-TR-0008-UA-RSCD byt Cobert et al). This effect is due to the field effect transistors (FETs) in the MIRI readout electronics  which do not instantaneously reset the detector signal, instead the exponential adjustment of the FETs after a reset causes the initial frames in an integration to be offset from their expected values.  As the RSCD effect is highly flux dependent, it would affect the shortest wavelengths far stronger than the longer wavelengths which see lower signal levels. The RSCD effect mainly influences the signal measured from the first few frames of an detector ramp implying that especially short integrations, like we have in the case of the L~168-9~b observations, will be affected by it. This might also explain why our noise estimate based on the in-orbit measured \texttt{PHOTOM} file does not show this increased noise behaviour towards shorter wavelengths. As the observations used to derive the \texttt{PHOTOM} file had 140 frames per integration, only a small percentage of the calibration data was effected by possible RSCD effects. This in contrast to the L~168-9~b data set with 9 frames per integration.


Figure  \ref{fig:fig_error_analysis} also shows that the noise measured in the \texttt{CASCADe} analysis exceeds estimates from simulations at $\lambda \gtrsim 11 \mu$m.  We speculate that this could be due to a difference in the applied spectral extraction methods. The \texttt{CASCADe} analysis used a fixed width extraction aperture while the \texttt{Eureka!} analysis applied a PSF weighted method. The latter method potentially results in higher signal to noise spectra for very faint signals at or below the background level, which we have at wavelengths $\lambda \gtrsim 11 \mu$m.


\section{Summary and Conclusions}\label{sec:conclusions}

We conducted an observation of the transiting planet L~168-9~b during JWST commissioning to assess the performance of the MIRI LRS mode in time series observations. This planet was selected to provide high spectro-photometric precision and to have a negligible transmission spectrum signal. This observation led to the following results:

\begin{enumerate}
    \item We analyzed the data using the first stage (Detector~1) of the STScI \texttt{jwst} pipeline, followed by independent spectral analyses using the \texttt{CASCADe} and \texttt{Eureka!} transit spectroscopy analysis packages. Both produced similar white-light curves, transmission spectra, and noise estimates. 
    
    \item \added{From the white light curve, we were able to reproduce the planet's transit depth to within the 1-$\sigma$ of the recent value compiled from all TESS data by \citet{patel2022}. The initial signal amplitude was approximately 0.25\% higher than the mean post-transit signal, and the excess signal decayed by a factor of $e$ within about 20 minutes. We have mitigated possible effects of this initial drift on the lightcurve fitting by either omitting the first 30 minutes of the timeseries or by explicitly fitting it. We fit and removed this detector signal settling with a two-component exponential and used the corrected light curve to refine the L~168-9 system's astrophysical parameters.}
    
    \item \added{The noise estimates in the transmission spectrum are approximately 20\% higher than predicted by random-noise simulations over wavelengths $6.8 \lesssim \lambda \lesssim 11$ $\mu$m with the adopted bin width of 0.149 $\mu$m. For wavelengths shorter than $\sim$7 $\mu$m we find significantly higher noise than predicted, up to a factor of 2 at the shortest wavelengths, which might be due to one or more detector effects affecting observation on very bright sources with relatively short detector ramps. The deviations seen at the longer than $\sim$11 $\mu$m between the 2 analysis methods could be due to differences in the applied spectral extraction methods. }
    
\end{enumerate}

\begin{acknowledgments}

We thank all people across the world who contributed to JWST's fantastic initial success; and in particular the entire MIRI commissioning team, the STScI cross-instrument TSO working group for JWST, the mission commissioning leads and project scientists. TPG and TJB acknowledge support from the NASA JWST project in WBS 411672.04.01.02. JB acknowledges support from the European Research Council under the European Union's Horizon 2020 research and innovation program ExoplANETS-A (GA No.~776403).

\end{acknowledgments}


\bibliography{miri_tso}{}
\bibliographystyle{aasjournal}

\facility{JWST (MIRI)}

\software{
astraeus \citep{astraeus2022},
astropy \citep{astropy_v5},
batman \citep{batman2015},
{CASCADe} \citep{2021A&A...646A.168C},
crds \citep{crds2022},
dynesty \citep{dynesty_v1_2_3},
{Exotethys} \citep{2020AJ....159...75M},
{Eureka!} \citep{eureka2022},
h5py \citep{h5py_v3-1-0},
jwst \citep{jwst2022},
matplotlib \citep{matplotlib2007},
numpy \citep{numpy2020},
numba \citep{10.1145/2833157.2833162},
pandas \citep{pandas_v1-4-3},
ray \citep{2017arXiv171205889M},
scipy \citep{scipy2020},
xarray \citep{xarray2017},
}

\appendix
\renewcommand\thefigure{A\arabic{figure}}
\setcounter{figure}{0}
\renewcommand\thetable{A\arabic{table}}
\setcounter{table}{0}

\section{\texttt{CASCADe} Analysis Additional Material}

\begin{figure*}[!htb]
\centering
\includegraphics[width=0.45\linewidth]{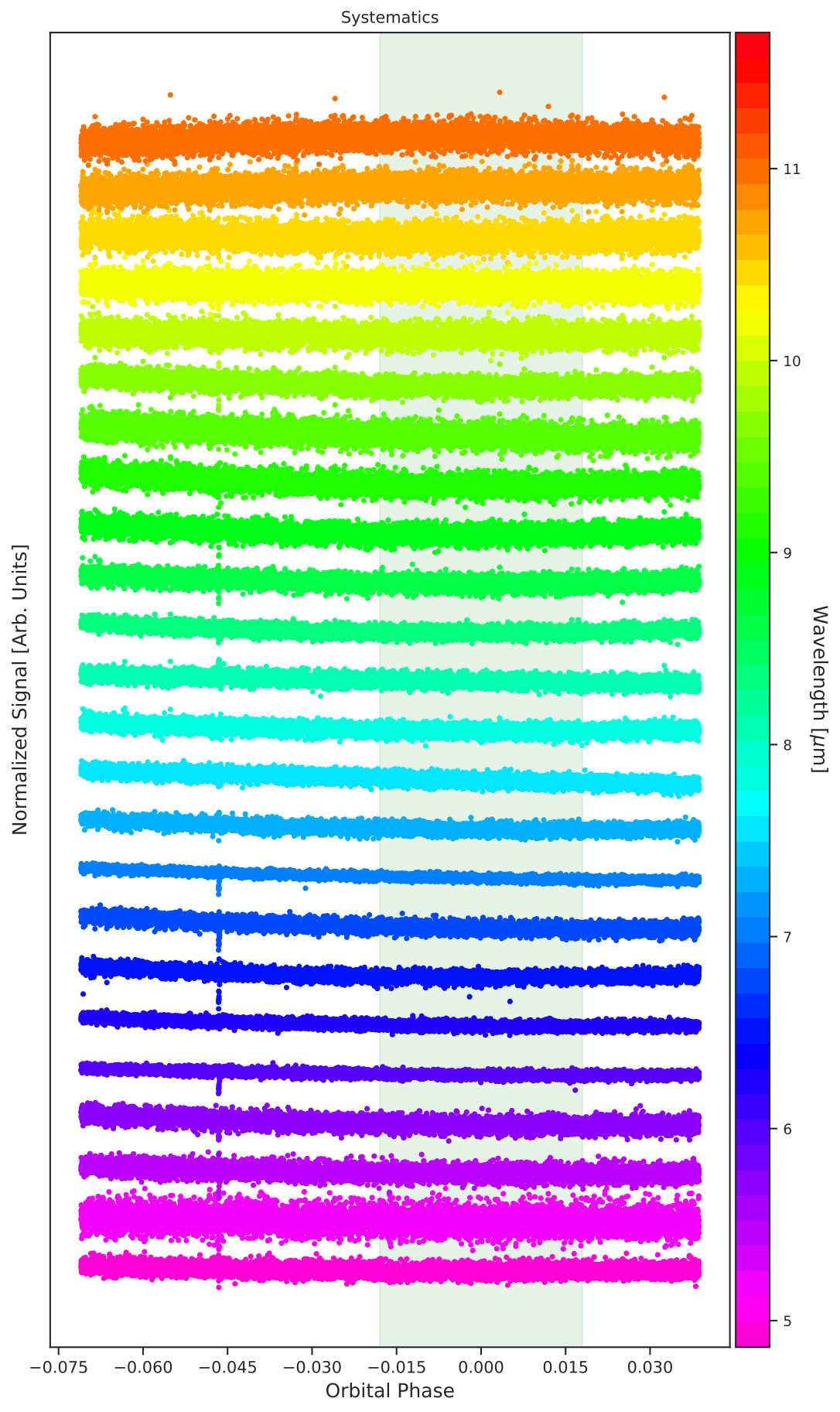}
\includegraphics[width=0.45\linewidth]{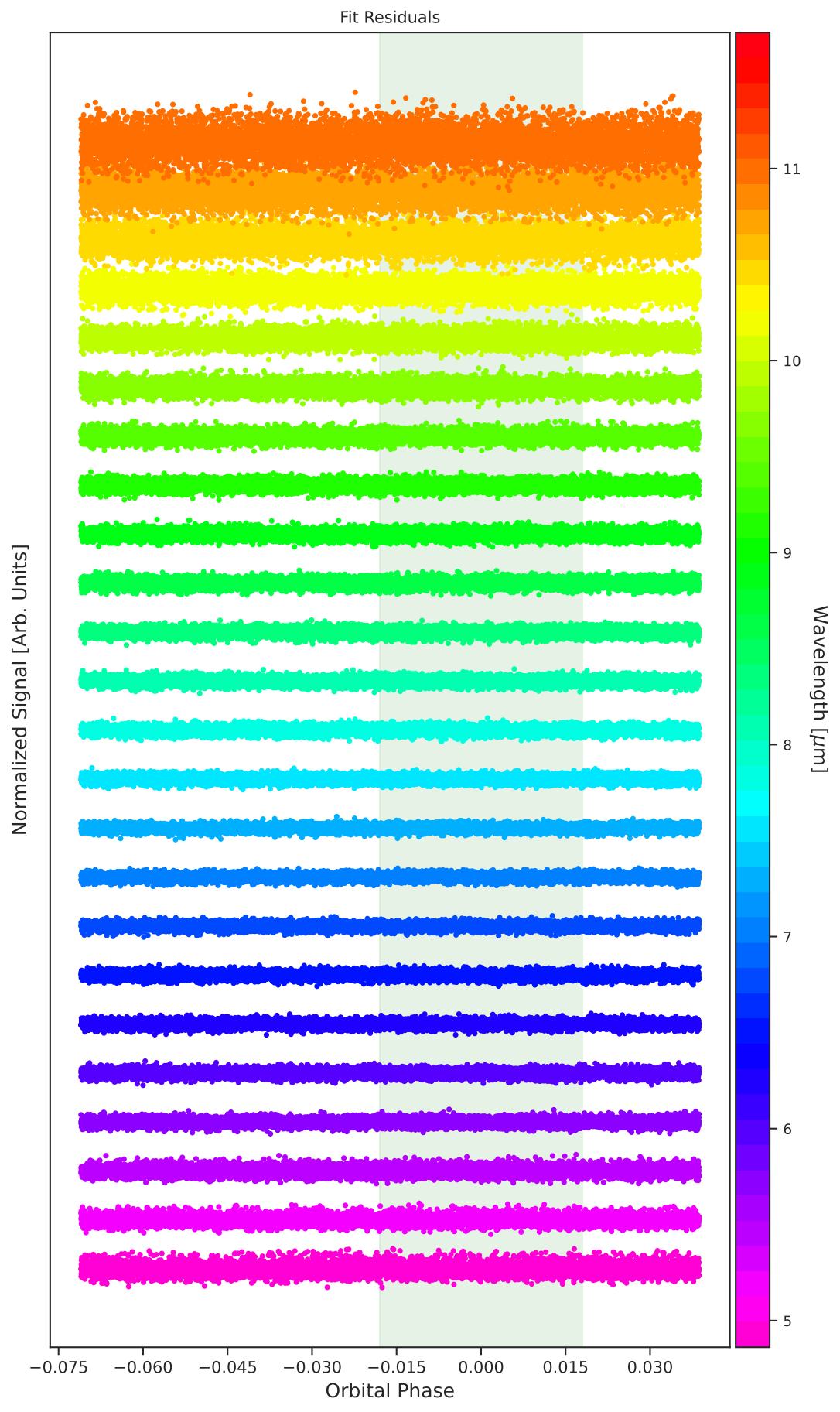}
\caption{\texttt{CASCADe}'s systematics model per wavelength bin (left) and residuals from the lightcurve fitting (right).
\label{fig:systematics_per_wavelength_and_residuals}}
\end{figure*}


\end{document}